\documentclass[showpacs,twocolumn]{revtex4}
\usepackage{amsmath}
\usepackage{graphics}
\begin{document}

\title{
Linear and nonlinear marginal stability for fronts of hyperbolic reaction diffusion equations}
\author{R.\ D.\ Benguria} 
\author{ M.\ C.\ Depassier}
\affiliation{
 Facultad de F\'\i sica\\
	Pontificia Universidad Cat\'olica de Chile\\
	       Casilla 306, Santiago 22, Chile}

\date{\today}
%\maketitle
\begin{abstract}
We study travelling fronts of equations of the form $u_{tt} + \phi(u) u_x = u_{xx} + f(u)$. A criterion for the transition from linear to nonlinear marginal stability is established for positive functions $\phi(u)$ and for any reaction term $f(u)$ for which the usual parabolic reaction diffusion equation $u_t = u_{xx} + f(u)$ admits a front. As an application, we treat reaction diffusion systems with transport memory.
\end{abstract}

\maketitle

\section{Introduction}
Reaction diffusion equations are used to model transport phenomena in a variety of contexts such as population dynamics, transmission lines, flame propagation among others. The prototype of such equations, for which a thorough understanding \cite{KPP37,AW78} of its properties has been achieved is the parabolic reaction diffusion equation
\begin{equation}
u_t = u_{xx} + f(u), \quad \text{with}\quad f(0) = f(1) = 0,
\label{parab}
\end{equation}
where subscripts denote derivatives. For positive reaction terms, sufficiently localized initial conditions evolve into a monotonic decaying travelling front joining the stable $u=1$ to the unstable $u=0$ equilibrium points. For bistable reaction terms,  which satisfy $f < 0$ in $(0,a)$ and $f > 0$ in $(a,1)$, with $\int_0^1 f >0$, it is possible to find initial conditions for which the system will evolve into a monotonic decaying front joining the two stable equilibrium points $u=1$ to $u=0$. In the first case there is a continuum of speeds for which there exist monotonic fronts, the system evolves into the front of minimal speed. In the second case there is a unique speed. 

The use of Eq. (\ref{parab}) to model a physical process involves assumptions on the stochastic process that describes the motion of the individuals. More specifically, Brownian motion is assumed \cite{mackean}. If instead a more realistic process is considered \cite{Dunbar},  then, in one spatial dimension, the differential equation that describes the motion is a hyperbolic reaction diffusion equation of the form
\begin{equation}
u_{tt} + \phi(u) u_t = u_{xx} + f(u).
\end{equation}
It has been shown that, as in the usual reaction diffusion equation, for positive $\phi(u)$, the hyperbolic equation admits monotonic decaying fronts of speed $c_h < 1$  and that the speed of the front is determined from a related parabolic equation \cite{Hadeler2}. 

Equations of this form have been studied for particular cases of $\phi(u)$ obtained assuming specific functions for memory effects in the diffusion term \cite{Mendez,Manne,Sancho,Abramson}. Recent work has dealt with the stability of the traveling front \cite{Gallay}.  For constant $\phi(u)$ and positive, concave, $f$ the traveling fronts are stable to local perturbations. The asymptotic behavior of the front of minimal speed has also been found \cite{Gallay}. Numerical solutions for positive but non concave $f$ indicate that the transition from linear to nonlinear marginal stability occurs at the same parameter values than for the parabolic equation Eq. (\ref{parab})\cite{Sancho}. Detailed analysis for a piecewise linear reaction term has given insight into the nature of possible travelling waves \cite{Manne}.  In systems with exponential transport memory  $\phi(u)$ is of the form 
$$
\phi(u ) = s - f'(u).
$$
For this case a variational principle for the front of minimal speed was constructed for positive reaction terms $f(u)$;  it was proved that  for concave reaction terms, as in the parabolic equation,  linear marginal stability holds \cite{Mendez}.

In this work we construct a variational principle for any positive function $\phi(u)$ and general reaction terms for which the parabolic equation admits a front. From this principle upper and lower bounds are constructed which permit the a priori determination of the transition from linear to nonlinear marginal stability \cite{VS88,VS89}. We recover all known results as particular examples,  and construct others that show different types of behavior depending on the explicit functions $f(u)$ and $\phi(u)$. Our main result is the following, consider Eq. (2), with $\phi(u) > 0$, and $f(u)$ a reaction term for which fronts of the parabolic equation exist, as described above. The minimal (or unique, for the bistable case) speed of the fronts joining the stable to the unstable points $u = 1$ to $u=0$ is given by
\begin{equation}
\frac{c^2}{1 -c^2} = \sup_g 2 \alpha \frac{ \int_0^1 g(K(u)) f(u) d\,u}
{\int_0^1 - g^2(u)/g'(u) d\,u}
\label{simpler}
\end{equation}
where
\begin{equation}
\frac{1}{\alpha} = \int_0^1 \phi(u) d\, u
\label{alpha}
\end{equation}
and
\begin{equation}
 K(u) = \alpha \int_0^u \phi(u) d \,u.
\label{K}
\end{equation}
The supremum is taken over all positive decaying functions $g(u)$ for which the integrals exist.
From this result it follows that, for positive reaction terms  which satisfy $f'(0) > 0$, the following bound holds,
$$
4 \frac{f'(0)}{ \phi^2(0)} \le \frac{c^2}{1 -c^2} \le 4 \sup_u \frac{\alpha f(u)}{\phi(u) K(u)}
\label{bound1}
$$
which enables one to characterize functions for which marginal stability holds. The lower bound, the marginal stability value,  is that imposed by linear considerations alone. Improved lower bounds valid for all reaction terms are obtained by direct use of the variational principle.

\section{Speed of fronts}

Consider the hyperbolic reaction diffusion equation
\begin{equation}
u_{tt} + \phi(u) u_t = u_{xx} + f(u),
\label{hypereq}
\end{equation}
with
$$ 
 \phi(u) >0  \qquad \text{and} \qquad f(0) = f(1) = 0,
$$
where $f(u)$ and $\phi(u)$ $\in C^1 [0,1]$. We shall assume that $f$ belongs to the class for which monotonic fronts joining the equilibrium points $u=1$ to $u=0$ exist. The precise conditions have been spelled above in the introduction, and in more precisely elsewhere \cite{AW78,frentes}. 

We wish to find the minimal or unique speed for  which there is a  monotonic decaying travelling wave solution 
$u = u(x-c^*\,t)$ of 
Eq.(\ref{hypereq}). The speed satisfies $(1 - c^2) u_{zz} + c \phi(u) u_z + f(u) = 0$,
$\lim u_{z \rightarrow -\infty} = 1$, $\lim u_{z \rightarrow \infty} = 0$, $u_z < 0$, where
$z = x - c t$. 
It is known that when $f>0$ there is a continuum of fronts for a range of speeds
$c^* < c < 1$, where $c^*$ is the minimal speed of a related parabolic equation \cite{Hadeler2}. 
There is a unique speed in the bistable case. We may, as usual, consider the trajectory in phase space by defining $p = - u_z (u)$. The monotonic decaying front obeys
\begin{subequations}
\label{phase}
\begin{eqnarray}
(1 - c^2) p \frac{d p}{d u} - c \phi(u) p+ f(u) = 0,  \\
\nonumber \\
p(0) = p(1) = 0, \qquad p >0 \quad  \text{in}\quad (0,1).
\end{eqnarray}
\end{subequations}
Before going any further we recall the constraints posed by linearization around the fixed points. Linearizing around $u=0$ we find that  $p$ approaches zero as 
$m u$, where $m$ is the positive root of
$$
(1 -c^2) m^2 - c \phi(0) m + f'(0) = 0,
$$
that is, 
\begin{equation}
m = \frac{1}{2} \frac{c \phi(0)}{1 -c^2}  + \frac{1}{2(1-c^2)} \sqrt{ c^2 \phi^2(0) - \frac{4 f'(0)}{1 -c^2}}.
\end{equation}
When $f'(0) > 0$, since it is known that monotonic fronts exist with $c^2 < 1$, $m$ is real if the term in square brackets is positive, that is,
\begin{equation}
\frac{c^2}{1 - c^2} \ge \frac{4 f'(0)}{\phi^2(0)} \equiv \frac{c_L^2}{1 - c_L^2},
\label{linearbound}
\end{equation}
where we have called $c_L$ the minimal speed derived from linear theory or, as it is known, linear marginal stability value.

When $f'(0) \le 0$, linear theory imposes no constraint on the speed. Analysis near the equilibrium point $u=1$ imposes no constarint on the speed as we assume that $f$ is positive or of the bistable type.

The simplest, but not unique, method to obtain a variational characterization of the speed is to introduce  a stretching of coordinates which reduces the equation to the standard parabolic reaction diffusion equation \cite{Hadeler2}. Since $\phi(u) > 0$, we may introduce a new independent coordinate defined by the following transformation,
\begin{equation}
y = K(u) 
\end{equation}
with $K(u)$ defined in Eq.(\ref{K}).
The transformation is invertible, and $y$ varies between 0 and 1. In the new coordinates, Eq.(\ref{phase}) reads
\begin{equation}
p(y) \frac{dp}{dy} - \frac{c}{\alpha (1 -c^2) } p(y) + \frac{f(K^{-1}(y))}{\alpha ( 1 -c^2) \phi(K^{-1}(y))} = 0,
\end{equation}
$$ p(0) = p(1) = 0,   \qquad p > 0 \quad \text{in}\quad (0,1).
$$
This equation is the equation for fronts of the parabolic equation, of speed
$$
\hat c = \frac{c}{\alpha ( 1 - c^2)}$$
and a reaction term
$$ 
F(y) = \frac{f(K^{-1}(y))}{\alpha ( 1 -c^2) \phi(K^{-1}(y))}.
$$
We now check that $F(y)$ satisfies all the requirements of the existence of fronts, if $f$ does. 
Effectively, $F(0) = F(1) =0$;  for $\phi > 0$ and $1 -c^2 >0$, sign(F) = sign(f), and 
$$
\int_0^1 F(y) d y = \frac{1}{1-c^2} \int_0^1 f(u) d u.
$$
So that if $f$ is of the bistable type, for which fronts exist, the same holds for $F$ if $1 - c^2 >0$ which we know to hold. Moreover 
$$
F'(y=0) = \frac{1}{\alpha^2 (1 - c^2)} f'(0).
$$
We may then apply directly the variational principle for fronts of the parabolic reaction diffusion equation \cite{frentes}, that is,
\begin{equation}
\hat c^2 = \max_g \frac{ 2 \int_0^1 g (y) F(y) d y}{\int_0^1 (-g^2(y)/g'(y)) d y}
\label{varhat}
\end{equation}
where the maximum is taken over all positive decaying functions $g$ in (0,1) for which the integrals exist. The maximum is attained for 
\begin{equation}
g = \hat g = \exp \left(-\int_{y_0}^y \frac{\hat c}{p(s)} d s\right), \qquad 0 < y_0 < 1
\label{ghat}
\end{equation}
provided that $F'(0) \le 0$, or, if $F'(0) > 0$, when $\hat m/\hat c > 1/2$. This is, 
$$
\frac{\hat m}{\hat c}  > \frac{1}{2} [ 1 + \sqrt{\hat c^2 - 4 F'(0)}\, ].
$$
Written in terms of the original quantities, $c, f, \phi$, this condition reads
$$
\frac{c^2}{1 - c^2} > \frac{ 4 f'(0)}{\phi^2(0)}.
$$
As in the usual parabolic case the maximum is attained except at the linear marginal stability value. Here too one can show that the linear marginal stability value is obtained by taking the supremum instead of the maximum.   In terms of the original functions $f$ and $\phi$ , Eq.(\ref{varhat}) reads,
\begin{equation}
\frac{c^2}{1 -c^2} = \max_g \frac {2 \alpha \int_0^1 (g (y) f(K^{-1}(y))/ \phi(K^{-1}(y))) d y}
{\int_0^1 (-g^2(y)/g'(y)) d y}.\label{var}
\end{equation}
Notice that the simpler form Eq.(\ref{simpler}) is obtained by changing to the original independent variable $u$.

\section{Criterion for Linear and Nonlinear Marginal Stability}

For the parabolic reaction diffusion equation Eq(\ref{parab}), for $f >0$, we know that the minimal speed, $c_{PF}$ of the propagating front is bounded below and above. When these two bounds coincide, then one can determine the speed unambiguosly.
A first estimate comes from the bound \cite{KPP37,AW78}
$$
2 \sqrt{f'(0)} \le  c_{PF} \le 2 \sup_u \sqrt{\frac{f(u)}{u}}.
$$
When these two bounds coincide linear marginal stability holds. It may still hold if these two bounds do not coincide. This can be decided making use of the integral variational principle which improves the lower bound \cite{frentes}, and of a minimax variational principle which improves the upper bound \cite{HR75}. Combined use of the two permits the exact determination of the speed. 
For the bistable case, we can obtain the speed making use of the integral variational principle. In this section the analogue of the above results is obtained for the hyperbolic fronts. 

\subsection{Upper bound}

We present here a 
simple derivation of the upper bound, valid both  for positive and bistable reaction terms. 
The following evident inequality holds,
\begin{equation*}
\frac{c^2}{1 -c^2} 
= \max_g \frac 
{2 \alpha \int_0^1 \frac{ g (y) f(K^{-1}(y))}{\phi(K^{-1}(y))} d y}
{\int_0^1 (-g^2(y)/g'(y)) d y} 
\end{equation*}

\begin{equation*}
\le \max_g \sup_y \left[ \frac {2 \alpha  f(K^{-1}(y))}{y \phi(K^{-1}(y))} \right]
\frac {\int_0^1 y g(y) dy}{\int_0^1 - g^2(y)/g'(y) dy}.
\end{equation*}
But, as shown in the Appendix,  
$$
\int_0^1 y g(y) d y \le 2 \int_0^1 - g^2(y)/g'(y) dy,
$$
from where we obtain
\begin{equation}
\frac{c^2}{1 -c^2} \le \sup_y \left[ \frac {4 \alpha  f(K^{-1}(y))}{y \phi(K^{-1}(y))} \right].
\end{equation}
Going back to the original independent variable, this is
\begin{equation}
\frac{c^2}{1 -c^2} \le \sup_u \left[ \frac {4 \alpha  f(u)}{K(u) \phi(u)} \right].
\label{upbound}
\end{equation}
Observe that this upper bound coincides with the linear value when the supremum occurs at $u=0$.

\subsection{Lower Bounds}

The lower bounds have already been obtained, one is the linear bound Eq.(\ref{linearbound}), and improved lower bounds can be obtained by choosing particular trial functions $g$. Equality, in the expression below, holds for a certain $g$.
\begin{equation}
\frac{c^2}{1 -c^2} \ge \max \left( \frac{c_L^2}{1 -c_L^2}, \frac 
{2 \alpha \int_0^1 \frac{ g (y) f(K^{-1}(y))}{\phi(K^{-1}(y))} d y}
{\int_0^1 (-g^2(y)/g'(y)) d y} \right).
\end{equation}

\section{Applications}

\subsection{Constant $\phi(u)$}
As a first example consider the case  $\phi(u) = \phi_0$, a constant. As explained in the introduction the stability of the traveling front for this equation has been proven recently \cite{Gallay}. Numerical investigations have shown that the transition from linear to nonlinear marginal stability occur at the same parameter values as for the parabolic equation with the same reaction term \cite{Sancho}. This last fact follows directly from the bounds above. For a constant $\phi(u)$, we obtain $\alpha = 1/\phi_0$ and $y = u$. Then, the bounds read
$$
4 f'(0) \le \frac{ \phi_0^2 c^2}{1-c^2} \le 4 \sup_u \frac{f(u)}{u}.
$$
The upper and lower bounds are identical to those for the speed of fronts of the parabolic equation parabolic equation; 
the speed is given by
\begin{equation}
 \frac{ \phi_0^2 c^2}{1-c^2} =
 \max_g 
\frac {2 \int_0^1 g(u) f(u) du}{\int_0^1 (-g^2(u)/g'(u)) d u} = C_{PF}^2,
\label{constfi1}
\end{equation}
The speed of the front is determined in terms of the speed of the parabolic front with the same reaction term, therefore the transition from linear to nonlinear marginal stability occurs at the same parameter values as for the parabolic case. The speed itself, however, is lower than the speed of the front of the parabolic equation,
\begin{equation}
c^2 = \frac{c_{PF}^2}{\phi_0^2 + c_{PF}^2} < c_{PF}^2.
\label{constfi2}
\end{equation}

\subsection{Suppresion of nonlinear marginal stability}
As a second example, which shows that substantially different behavior may occur, take the reaction function $f(u) = u ( 1 - u) ( 1 + a u)$. The front of minimal speed for the parabolic equation is
given by
\begin{eqnarray*}
C_{PF} &=& 2 \qquad \text{for}\quad 0 < a  < 2 \\
C_{PF} &=& \sqrt{\frac{2}{a}} + \sqrt{\frac{a}{2}} \qquad \text{for} \quad a > 2.
\end{eqnarray*}
If we now consider the hyperbolic equation with $\phi(u) = 1 + a u$, the upper and lower bounds coincide and the speed 
$$
\frac{c^2}{1-c^2} = 4 \qquad \text{for all}\quad a.
$$
Effectively we obtain 
$$
4 = 4 f'(0) \le \frac{c^2}{1-c^2} \le 4 \sup_u \frac{1-u}{1 + a u/2} = 4.
$$
The minimal speed is the linear marginal stability value, no transition from linear to nonlinear marginal stability occurs. A significant slowdown of the speed takes place. 

\subsection{Time delayed diffusion}

In systems with time delayed diffusion the function $\phi(u)$ adopts the form
$$
\phi(u) = s - f'(u)
$$
form which arises when considering diffusion with exponential time delay. Here $s$ is a parameter.
For this function $\alpha = 1/s$ and $K(u) = u - f(u)/s$.
The speed of the front was studied for positive reaction terms, and it was shown that for concave reaction terms marginal stability holds \cite{Mendez}. 
This can be proved easily from the present approach. As Eq.(\ref{upbound}) shows, linear marginal stability holds for all positive reaction terms with $f'(0) > 0$ provided that the supremum of 
\begin{equation}
B(u) =  \frac {4 \alpha  f(u)}{K(u) \phi(u)}
\end{equation}
occurs on $u=0$, since $B(0) = 4 f'(0)/\phi^2(0)$, is equal to the lower bound.
To prove then that linear marginal stability holds it suffices to show that $B'(u) < 0$ for $u \in (0,1]$. Taking the derivative we obtain, after some algebra, 
$$
B'(u) = 4 \frac{\alpha}{K(u)} \left[ \frac{f f''}{\phi^2(u)}+ \frac{s \alpha}{\phi(u) K(u)} h(u)\right]
$$
where $h(u) = u f' - f$. It follows immediately that for concave reaction terms $B'(u) < 0$. For concave terms $ f'' < 0$, so the first term in the square brackets is negative. And $h(u)$ is negative for concave functions as well, $h(0) = 0$, and $ h' = u f'' < 0$.  So we have recovered the result that for concave functions with $f'(0) > 0$, linear marginal stability holds. As it occurs for the parabolic case. it may hold for noncave functions as well, fact which can be determined by use of the variational principle. 

Consider again the reation term $f(u) = u (1-u) (1 + a u)$ as in subsection {\bf B}. In the parabolic equation transition from linear to nonlinear marginal stability occurs at $a=2$. 
Choose here $s=2$ which guarantees $\phi> 0$ for $0 < a < (5 + \sqrt{21})/2 \approx 4.79$ . We show in Fig. 1 that the transition from linear to nonlinear marginal stability occurs at a lower value of $a$ than that for the parabolic case. We chose as a trial function $g(u) = (1-u)^7 / u^{1.8}$. The dot-dash line shows the lower bound imposed by the linear marginal stability value, the solid line the lower bound from the variational principle. The transition occurs at at $a$ as low as 1.6 at least.  The precise transition value  can be found by using a better trial function. The dashed line shows the upper bound, which guarantees that linear marginal stability holds for $0 < a < 1$, values for which $f$ is concave. 

\subsection{A bistable reaction term}

The variational principle and the bounds obtained above hold for positive and bistable reaction systems, no assumption on the sign of $f$ is made in the derivations. We may therefore apply them to a bistable reaction term. We take as an example a case for the speed of the parabolic front can be determined exactly, 
$$
f( u) = u (1 - u) (u - a) \qquad \text{for}\quad 0 < a < 1/2
$$
The unique value of the speed for which a front exists is
$$
c_{PF} = \frac{1}{\sqrt{2}} - a \sqrt{2}
$$
 If $\phi(u) = \phi_0$ a constant, the speed of the front is given by Eq(\ref{constfi2}), and the exact speed is known with $c_{PF}$ given above. 

For this bistable reaction term, with $\phi(u) = 1 - f'(u)$, which is positive for $0 < a < 1/2$, with a simple trial function $g(u) = 1 - u$, all integrals are elementary, we obtain, using Eq.(\ref{var}),
$$
c^2 \ge \frac{18 -49 a + 14 a^2}{88- 49 a + 14 a^2}.
$$
Numerical determination of the speed can be made with any desired accuracy by use of improved trial functions.

We wish to point that,  in recent work,  a variational principle valid only for positive reaction terms has been applied to a bistable example \cite{Mendez}. Bistable reaction terms, by definition, are not positive throughout the whole interval, so the hypothesis of their derivation is violated. We attribute the nearly perfect agreement  between  the speed of the front obtained from their (nonapplicable) variational principle to that obtained from direct numerical integration to a coincidence for the specific trial function chosen. It is not difficult to find an acceptable trial function, sharply peaked at the origin, for which the integrand of their variational expression, and so the speed, becomes complex.

\section{Conclusion}

We have studied the speed of travelling fronts of hyperbolic reaction diffusion equations (\ref{hypereq}) for reaction terms for which the parabolic reaction diffusion equation (\ref{parab}) admits fronts. 
A variational principle for the speed of the fronts has been obtained for a wide class of systems. Explicit upper and lower bounds, obtained from the variational expression allow one to characterize sytems for which marginal stability holds. We have seen that depending on the nonlinearities of the equation the range of validity of linear marginal stability, can be increased or decreased, compared to the comparable range for a parabolic equation with the same reaction term. The stability of the fronts has been proved for a limited class of such equations \cite{Gallay}, numerical solutions have been explored for systems with memory effect. These results will be more relevant if the stability of fronts of more general systems is proved. At present numerical results indicate that the front of minimal or unique speed can be selected. 
The general initial value problem for hyperbolic reaction diffusion equations is an open problem.

\appendix
\section{}

Let $g(y)$ be a positive decaying function, which satisfies $g(1) =0$, and call $h(y)= -g'(y)$, so that $h >0$. 
Then
\begin{eqnarray*}
\left[ \int _0^1 y g(y) du \right]^2  &=& \left[ \int _0^1 \left(\frac{g}{\sqrt{h}}\right) (\sqrt{h} y) du \right]^2 \\
&\le&   \int_0^1 \frac{g^2(y)}{h(y) } d y \int_0^1 y^2 h(y) d y,
\end{eqnarray*}
where we used Schwarz's inequality. The second integral on the right side is, integrating by parts,
$$
\int_0^1 y^2 h(y) d y = \int_0^1  y^2 (-g'(y)) d y = 2 \int_0^1 y g(y) d y,
$$
where we used the fact that $g$ diverges at $y=0$ slower than $(1/y^2)$ (this is equivalent to $m/c > 1/2$) and that $g(1) =0$ which can be seen form Eq.(\ref{ghat}).
Replacing this in the inequality above we have the desired result, 
$$
\int_0^1 y g(y) d y \le 2 \int_0^1 g^2(y)/h(y) dy.
$$

\begin{figure}[h]
\includegraphics{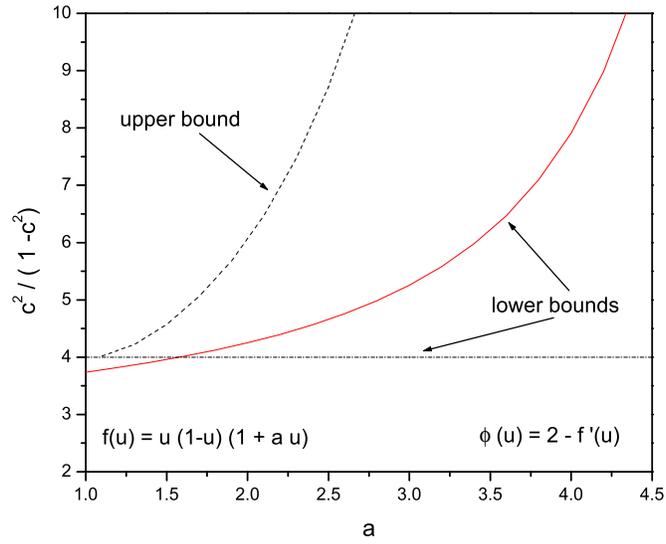}
\caption{Bounds on the speed of fronts for an example with time delayed diffusion. The range of validity of linear marginal stability is reduced compared to the parabolic diffusion equation
with the same reaction term.}
\end{figure}

 \end{document}